\documentclass{iopjournal}
\usepackage{graphicx, amsmath, amssymb, amsthm, hyperref, multirow, xcolor}
\usepackage[caption=false]{subfig}
\allowdisplaybreaks[1] 

\begin{document}

\articletype{Paper}


\newcommand{\comb}[2]{{\begin{pmatrix} #1 \\ #2 \end{pmatrix}}}
\newcommand{\braket}[2]{{\left\langle #1 \middle| #2 \right\rangle}}
\newcommand{\bra}[1]{{\left\langle #1 \right|}}
\newcommand{\ket}[1]{{\left| #1 \right\rangle}}
\newcommand{\ketbra}[2]{{\left| #1 \middle\rangle \middle \langle #2 \right|}}

\newcommand{\fref}[1]{Fig.~\ref{#1}}
\newcommand{\Fref}[1]{Figure~\ref{#1}}
\newcommand{\sref}[1]{Sec.~\ref{#1}}
\newcommand{\Sref}[1]{Section~\ref{#1}}
\newcommand{\subsref}[1]{Subsec.~\ref{#1}}
\newcommand{\tref}[1]{Table~\ref{#1}}


\title{Self-Trapping Bounds for Continuous-Time Nonlinear Quantum Walks on Path and Cycle Graphs}

\author{Yujia Shi$^{1,*}$\orcid{0000-0002-4044-5659}, Thomas G.~Wong$^{1,*}$\orcid{0000-0002-5019-7302}}

\affil{$^1$Department of Physics, Creighton University, 2500 California Plaza, Omaha, Nebraska 68178, USA}

\affil{$^*$Author to whom any correspondence should be addressed.}

\email{yujiashi@creighton.edu}
\email{thomaswong@creighton.edu}


\begin{abstract}
	We explore a continuous-time quantum walk starting at a single vertex on the discrete path and cycle with a cubic nonlinearity. Such nonlinearities arise in Bose-Einstein condensates described by the Gross-Pitaevskii equation or by nonlinear optical waveguide arrays. When the nonlinearity is sufficiently strong, the walker remains localized at its initial vertex, a phenomenon known as self-trapping. This contrasts with linear quantum walks, which are known for spreading quickly in one dimension. While self-trapping has been known numerically, we introduce an analytical method that proves self-trapping and yields a quantitative relationship between the nonlinearity coefficient and the trapping probability. We propose that this trapping can be used for timing in quantum state transfer, where a qubit is held at a node until it is ready to be transferred, and it can also be held again at the receiving node. This scheme can also be interpreted as a form of quantum memory, with the trap and transfer corresponding to the storage and release of quantum information.
\end{abstract}


\section{Introduction}

From the time researchers first considered quantum versions of classical random walks, many of their explorations centered around how quantum walks evolve in one dimension (1D) \cite{Aharonov1993,Meyer1996a,Meyer1996b,Nayak2000,Ambainis2001,benAvraham2004}. Various properties of quantum walks were first observed in 1D, such as the need for a discrete-time quantum walk to contain an internal degree of freedom \cite{Meyer1996b}, as well as quadratically faster spreading and mixing for quantum walks than random walks \cite{Ambainis2001}. This inspired additional investigation into other graphs, including quantum walks exhibiting exponentially faster hitting than random walks on the hypercube from one corner to the other \cite{Kempe2003,Kempe2005}, to an exponential speedup over all classical algorithms when traversing welded trees \cite{Childs2003}. Since then, quantum walks have been applied to a variety of problems, such as spatial searching \cite{SKW2003}, solving element distinctness \cite{Ambainis2004}, finding triangles \cite{MSS2005}, evaluating boolean formulas \cite{FGG2008}, and performing universal quantum computation \cite{Childs2009}.

Although Schr\"odinger's equation, the fundamental equation of quantum mechanics, is linear, there are systems whose effective evolutions are described by nonlinear Schr\"odinger-type equations, such as the following with a cubic nonlinearity:
\begin{equation}
\label{eq:NLSE}
i\hbar\,\frac{\partial \psi}{\partial t}
=
\bigl(H - g|\psi|^2\bigr)\psi,
\end{equation}
where $g$ is a real-valued coefficient of the nonlinearity $|\psi|^2\psi$. For example, in the mean-field limit, Bose-Einstein condensates are described by the Gross-Pitaevskii equation \cite{Dalfovo1999}, which has the form of this cubic nonlinear Schr\"odinger equation \eqref{eq:NLSE}, with $g$ proportional to the negative of the scattering length of the Bose-Einstein condensate, which can be arbitrarily tuned via Feshbach resonance \cite{Timmermans1999} by adjusting an external magnetic field. So, positive $g$ corresponds to negative scattering lengths, meaning the interactions between condensate interactions are attractive, while negative $g$ corresponds to positive scattering lengths and repulsive interactions. As another example, waveguide arrays in nonlinear optics behave according to \eqref{eq:NLSE} \cite{Christodoulides2003,Lederer2008}, and while $g$ can be dynamically adjusted \cite{Cui2022}, its range is more limited than that of Bose-Einstein condensates.

\begin{figure}
\begin{center}
    \subfloat[] {
        \includegraphics{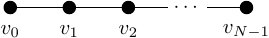}
        \label{fig:path-N}
    } \qquad\qquad\qquad
    \subfloat[] {
        \includegraphics{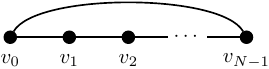}
        \label{fig:cycle-N}
    }
    \caption{\label{fig:1D}(a) The path graph of $N$ vertices, i.e., $P_N$, and (b) the cycle of $N$ vertices, i.e., $C_N$.}
\end{center}
\end{figure}

Utilizing this effectively nonlinear behavior for quantum computing, it has been shown that a continuous-time quantum walk can search more quickly when it evolves by the cubic nonlinear Schr\"odinger equation \cite{Wong3}, at the expense of the number of bosons needed for the effective nonlinearity to hold. But in this paper, we return to the seminal question of how a quantum walk evolves in 1D, but now with a nonlinear quantum walk. In particular, we explore how a continuous-time quantum walk governed by the cubic nonlinear Schr\"odinger equation \eqref{eq:NLSE} evolves on a 1D lattice of $N$ vertices with open boundaries (i.e., on a path graph) and with periodic boundaries (i.e., on a cycle graph), examples of which are shown in \fref{fig:1D}, from an initially localized point.

In the next section, we give the precise definition of the 1D nonlinear quantum walk and review previous numerical results demonstrating that when the nonlinearity coefficient $|g|$ is sufficiently large, the walker is trapped at its initial vertex, which is called self-trapping because it is induced by the nonlinear effects of the system itself. Then, in \sref{sec:analysis}, we analytically prove a relationship between $|g|$ and the self-trapping probability. Using it, given a value of $|g|$, we can find a lower bound on the self-trapping probability, and conversely, given a desired self-trapping probability, we can give a lower bound on the value of $|g|$ that is needed to achieve it. In \sref{sec:transfer}, we propose a potential application for timing quantum state transfer, which can also be considered a form of quantum memory. We conclude in \sref{sec:conclusion}.


\section{\label{sec:numerics}Self-Trapping in 1D Nonlinear Quantum Walks}

We label the $N$ vertices of the lattice $v_0, v_1, \dots, v_{N-1}$, as shown in \fref{fig:1D}, and denote their respective basis vectors $\ket{0}, \ket{1}, \dots, \ket{N-1}$. Then, the state of the system is described by a column vector $\ket{\psi(t)} \in \mathbb{C}^N$. The walker is initially localized at a single vertex $v_r$, so the initial state is
\[ \ket{\psi(0)} = \ket{r}. \]
We denote the $j$-th component of $\ket{\psi}$, or amplitude at vertex $v_j$, by $\psi_j(t) = \braket{j}{\psi(t)}$. Then, $|\psi_j(t)|^2$ is the probability of measuring the walker at vertex $v_j$ at time $t$.

To effect the quantum walk, the linear Hamiltonian $H$ in \eqref{eq:NLSE} is proportional to the (negative of the) adjacency matrix $A \in \mathbb{C}^{N \times N}$ of the graph \cite{Childs2003}, defined by $A_{ij} = 1$ if vertices $i$ and $j$ are adjacent, and $A_{ij} = 0$ otherwise:
\[ H = -\gamma A, \]
where $\gamma$ is the jumping rate of the quantum walk. Without loss of generality, we take $\gamma = 1$ throughout the paper \cite{FG1998b}. In our subsequent results, to restore $\gamma$, we can replace $g \to \gamma g$ and $t \to \gamma t$. We also take $\hbar = 1$. Then, according to the cubic nonlinear Schr\"odinger equation \eqref{eq:NLSE}, the $j$-th amplitude of $\ket{\psi(t)}$ evolves by
\[
i\,\frac{d \psi_j}{dt}
=
- \sum_{k=0}^{N-1} A_{jk}\,\psi_k
- g\,|\psi_j|^2\,\psi_j.
\]

This one-dimensional nonlinear quantum walk from a single vertex is equivalent, up to a rescaling of time, to previous work \cite{Kenkre1986,Molina1993} on the discrete nonlinear Schr\"odinger equation. These previous works were done in the context of polaritons in condensed matter physics and nonlinear waveguides in optical physics, rather than our motivation of quantum walks and quantum information science. Existing analytical results are limited to very small systems, including paths of length $2$ \cite{Kenkre1986} and cycles of length $3$ \cite{Molina1993}, whereas larger path and cycle graphs have previously only been investigated numerically \cite{Molina1993}. In particular, for general $N$, when $g$ is sufficiently large, the probability amplitude remains localized near the initial vertex \cite{Molina1993}. This self-trapping has even been experimentally demonstrated in Bose-Einstein condensates in optical lattices \cite{Anker2005} and nonlinear waveguide arrays \cite{Eisenberg1998}. Self-trapping has also been explored in discrete nonlinear Schr\"odinger equations in the presence of nonlinear impurities, including on infinite paths \cite{Chen1993}; paths of length 2 (i.e., the dimer) \cite{Bustamante1998}; and 1D, square, triangular, simple cubic, and Bethe lattices \cite{Bustamante2000}.

\begin{figure*}
    \includegraphics[width=\linewidth]{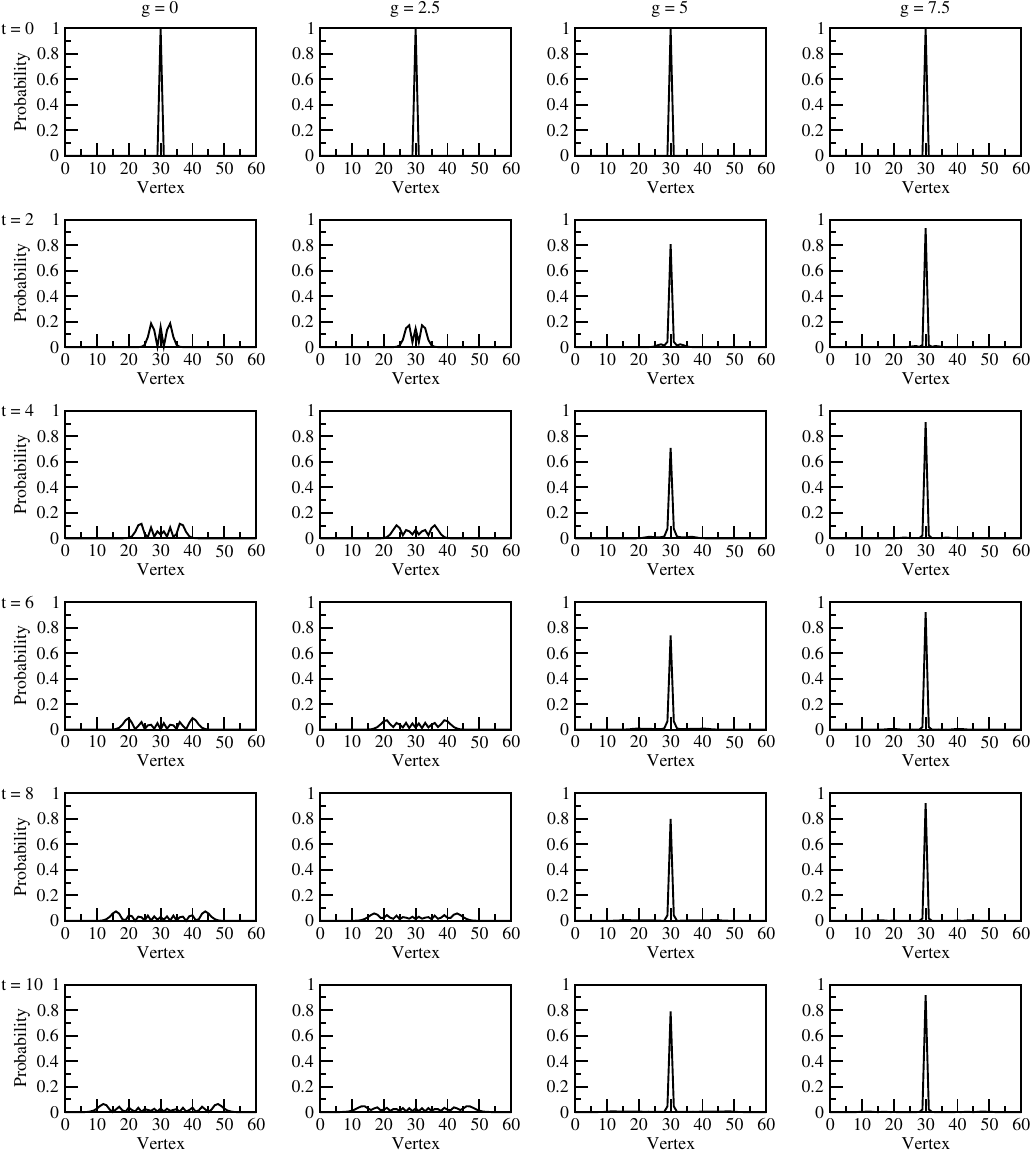}
    \caption{\label{fig:prob-dist}For the adjacency quantum walk on $P_{61}$ initially at the middle vertex (i.e., $v_{30}$), the probability distribution across all vertices at various times $t = 0, 2, 4, 6, 8, 10$ with various nonlinearity coefficients $g = 0, 2.5, 5, 7.5$.}
\end{figure*}

\begin{figure*}
    \includegraphics[width=\linewidth]{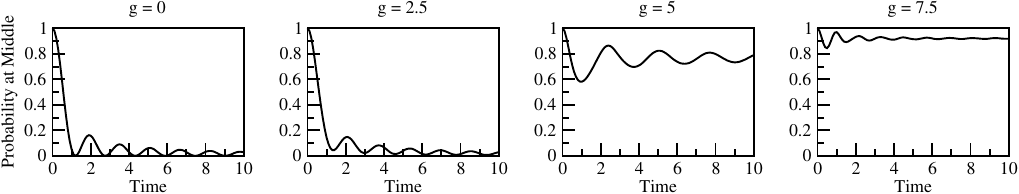}
    \caption{\label{fig:prob-initial}For the adjacency quantum walk on $P_{61}$ initially at the middle vertex (i.e., $v_{30}$), the probability at this middle vertex $|\psi_{30}(t)|^2$ as a function of time, for various nonlinearity coefficients $g = 0, 2.5, 5, 7.5$.}
\end{figure*}

As an example of self-trapping on a path \cite{Molina1993}, in \fref{fig:prob-dist}, we plot the probability distribution of the quantum walk at various times and with various $g$'s, when walking on $P_{61}$ with the walker initially at the middle vertex $v_{30}$. In the first column of plots, $g = 0$, which is the linear quantum walk. The top plot is the probability distribution at $t = 0$, and the walker is entirely at $v_{30}$. Going down, the subsequent plots show the probability distribution at $t = 2, 4, 6, 8, 10$, and as time progresses, the probability spreads out, further and further from the middle. The second column is $g = 2.5$, and with this small nonlinearity, the probability distribution evolves similarly to the linear ($g = 0$) case. The third column is $g = 5$, and now the nonlinearity is strong enough such that the evolution is very different. Now, the probability of staying at $v_{30}$ remains high throughout the evolution, which is the self-trapping observed by Molina \textit{et al.} The fourth column is $g = 7.5$, and the self-trapping is even stronger, with the probability of the walker staying at $v_{30}$ staying near 1. The absence and presence of self-trapping is also evident in \fref{fig:prob-initial}, where for the same path $P_{61}$ and parameters, we plot the probability at the middle vertex $v_{30}$ as a function of time. When $g = 0$ and $g = 2.5$, the probability at the middle vertex drops from 1 to near 0, whereas when $g = 5$ and $g = 7.5$, the probability is trapped, with larger $g$ resulting in more probability being trapped. The Mathematica code used to generate both \fref{fig:prob-dist} and \fref{fig:prob-initial} is available in \cite{data}.

\begin{figure}
\begin{center}
    \subfloat[] {
        \includegraphics[width=7.2cm]{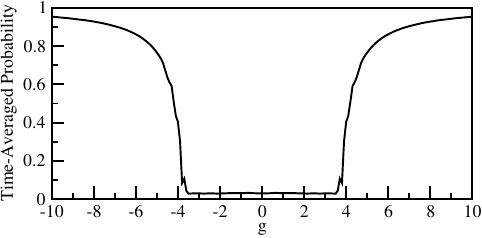}
        \label{fig:time-averaged-P61-mid}
    } \quad
    \subfloat[] {
        \includegraphics[width=7.2cm]{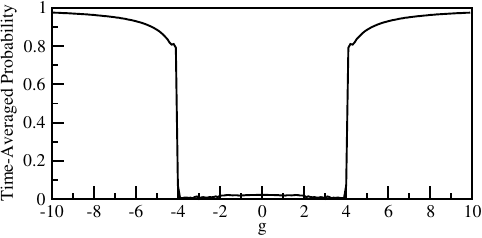}
        \label{fig:time-averaged-P61-end}
    }
    
    \subfloat[] {
        \includegraphics[width=7.2cm]{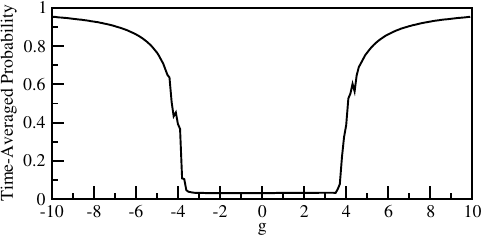}
        \label{fig:time-averaged-C61}
    }
    
    \caption{\label{fig:time-averaged}For adjacency quantum walks, the discrete approximation of the time-averaged probability $\bar{p}_r$ as a function of the nonlinearity coefficient $g$, using a total simulation time of $T = 300$ and $M = 400$ points for each. (a) is a quantum walk on $P_{61}$ initially at the middle vertex (i.e., $v_{30}$), (b) is a quantum walk on $P_{61}$ initially at an endpoint $v_0$, and (c) is a quantum walk on $C_{61}$ where the initial vertex does not matter.}
\end{center}
\end{figure}

To quantify this self-trapping, Molina et al.~\cite{Molina1993,Chen1993,Bustamante2000,Yue2014} considered the time-averaged probability at the initial vertex $v_r$:
\[ \lim_{T \to \infty} \frac{1}{T} \int_{0}^{T} |\psi_r(t)|^2\,dt. \]
Numerically, we calculate the probability at the initial vertex for finite time, such as in \fref{fig:prob-initial}, and the initial drop in probability can affect the average over a finite interval. So instead, we only average over the latter half of the simulation interval:
\[
\frac{2}{T}\int_{T/2}^{T} |\psi_{\mathrm{mid}}(t)|^2\,dt.
\]
Furthermore, this quantity is approximated by sampling the probability distribution over $M$ evenly spaced points in the second half of the simulation interval, resulting in the discrete average
\[
\bar{p}_r
=
\frac{1}{M} \sum_{j=1}^{M} |\psi_{\mathrm{mid}}(t_j)|^2.
\]
where $\{t_j\}_{j=1}^M$ are uniformly sampled points in the interval $\left[T/2,\,T\right]$. For example, for a quantum walk on $P_{61}$ initially at the midpoint $v_{30}$, this discrete approximation to the time-averaged probability $\bar{p}_\text{mid}$ is plotted in \fref{fig:time-averaged-P61-mid} for various values of $g$ using a simulation time of $T = 300$ and $M = 400$ samples. We see that for small $|g|$, the walker is not trapped, but for large $|g|$, the walker is trapped. With the walker is initially at an endpoint instead, we get the time-averaged probability shown in \fref{fig:time-averaged-P61-end}. Walking on the cycle, where it does not matter which vertex we start with, the time-averaged probability is shown in \fref{fig:time-averaged-C61}. In all three of these plots, the critical value of $g$, which separates the absence and presence of self-trapping, is $|g_c| = 4$, which is what a previous numerical study also found \cite{Molina1993}. Our \fref{fig:time-averaged} is similar to Fig.~3 and Fig.~4 of \cite{Molina1993}, including the jagged anomalies, which could be due to the number of vertices $N$, averaging time $T$, number of samples $M$, or numerical solver. The Mathematica code used to generate \fref{fig:time-averaged} is included in \cite{data}. Nonetheless, the general takeaway that a critical nonlinearity coefficient exists at $|g_c| = 4$ remains. As noted in the conclusion of \cite{Molina1993}, the asymmetry in \fref{fig:time-averaged-C61} is because of the odd number of vertices; with an even number, the time-averaged probability is symmetric under a change of sign in $g$.


\section{\label{sec:analysis}Proof of Self-Trapping}

As discussed in the last section, self-trapping on paths and cycles with general $N$ has been known numerically since at least the year 1993 \cite{Molina1993}. Over thirty years later, we now give the first analytical proof of it. We will derive a bound where, given $|g|$, we can determine that the self-trapping probability is at least some value, and conversely, given a desired self-trapping probability, we will prove that it is achievable if $|g|$ is at least some value. As a consequences of this method, with the quantum walk initially at an endpoint of a path, self-trapping occurs when $|g| \gtrapprox 7.22$, and for a quantum walk initially at an interior point of a path, or anywhere on a cycle, self-trapping occurs when $|g| > 9$. These are not tight bounds, as we just saw from that it should be $|g| > 4$. Nonetheless, these are the first rigorous bounds for this problem, and we hope the proof method may find application to other graphs or inspire refined methods that improve the tightness of the result.

\begin{figure}
\begin{center}
    \subfloat[$\deg(v_r) = 1$] {
        \includegraphics[width=7.2cm]{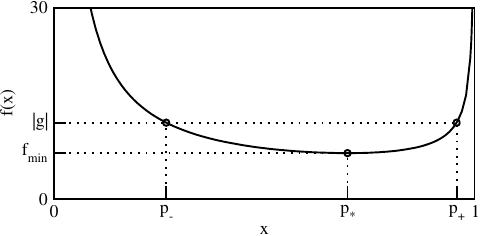}
        \label{fig:function-deg1}
    } \quad
    \subfloat[$\deg(v_r) = 2$] {
        \includegraphics[width=7.2cm]{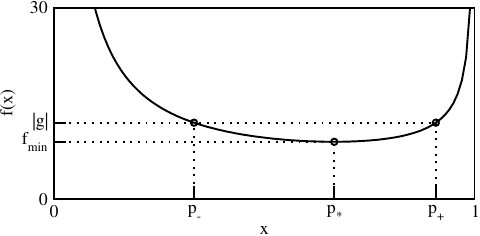}
        \label{fig:function-deg2}
    }
    \caption{\label{fig:function}A plot of the function $f(x)$ with (a) $\deg(v_r) = 1$ and (b) $\deg(v_r) = 2$. In both plots, the minimum of $f(x)$ is labeled, along with where it equals $|g| = 12$.}
\end{center}
\end{figure}

To begin, let us denote the self-trapping probability by
\[ p(t) = |\psi_r(t)|^2. \]
As proved in Appendix~\ref{appendix}, this self-trapping probability $p(t)$ evolves in such a way that
\begin{equation}
    \label{eq:bound-g-p}
\frac{2 \sqrt{\operatorname{deg}(v_r)}}{\sqrt{p(t)[1-p(t)]}}
+
\frac{2}{p(t)}
\ge
|g|
\end{equation}
for all $t$. Let us call the function on the left-hand side $f$:
\[
    f(x) =
\frac{2\sqrt{\operatorname{deg}(v_r)}}{\sqrt{x(1-x)}}
+
\frac{2}{x}.
\]
In \fref{fig:function-deg1}, we have plotted $f(x)$ with $\operatorname{deg}(v_r) = 1$, corresponding to a quantum walk on a path initially at an endpoint. Similarly, in \fref{fig:function-deg2}, we have plotted $f(x)$ with $\operatorname{deg}(v_r) = 2$, corresponding to a quantum walk on a path initially at an interior point, or a quantum walk on a cycle initially from any point. The second derivative of $f(x)$ is
\[ \frac{\mathrm{d}^2 f}{\mathrm{d}x^2} = \sqrt{\operatorname{deg}(v_r)} \frac{2x(1-x) + \frac{3}{2} (2x-1)^2}{[x(1-x)]^{5/2}} + \frac{4}{x^3}, \]
which is positive when $0 < x < 1$, and so $f(x)$ is concave up in the region. Furthermore, $f(x)$ is continuous in $(0,1)$ and diverges to $+\infty$ as $x \to 0^+$ and $x \to 1^-$, so it attains a global minimum at some $p_*\in(0,1)$:
\[
f_\text{min} = f(p_*).
\]
This minimum is identified in \fref{fig:function}. Now, consider some $|g|>f_\text{min}$, such as in \fref{fig:function}. Since $f(x)$ is concave up, the equation $f(x)=|g|$ has two solutions, one to the left of $p_*$ called $p_-$, and the other to the right called $p_+$, as shown in \fref{fig:function}. Between $p_-$ and $p_+$, $f(x)$ must be less than $|g|$, as shown in \fref{fig:function}. That is,
\[
f(x)<|g| \quad \text{for } x\in(p_-,p_+).
\]
But from \eqref{eq:bound-g-p}, $f(p(t)) \ge |g|$, so it follows that $p(t)$ cannot be between $p_-$ and $p_+$, i.e.,
\[
p(t)\notin(p_-,p_+)
\quad \text{for all } t \ge 0.
\]
Finally, since the walker is initially localized, $p(0) = 1$, and since $p(t)$ is continuous, it cannot drop below $p_+$ into the interval $(p_-,p_+)$. Therefore, $p(t)$ must stay above $p_+$ for all time, i.e.,
\[
p(t)\ge p_+
\quad \text{for all }t\ge0.
\]
Thus, we have proved that given $|g| > f_\text{min}$, self-trapping occurs with probability lower bounded by $p_+$, where $p_+$ is the larger solution to $f(x) = |g|$.

Note that $f(x)$ does not depend on the number of vertices $N$. It only depends on the degree of the initial vertex $\operatorname{deg}(v_r)$. With the initial vertex as an endpoint of a path, the minimum of $f(x)$ occurs at $p_* \approx 0.698$ with $f_\text{min} \approx 7.22$, and so our result says that self-trapping occurs when $|g| \gtrapprox 7.22$. With the walk initially at an interior vertex of a path, or any vertex on a cycle, $v_r$ has degree 2, and $f$ attains its minimum at $p_* = 2/3$ with a value of $f_\text{min} = 9$. Then, our result says that self-trapping occurs when $|g| > 9$. As discussed earlier, these are not tight bounds, as \fref{fig:time-averaged} shows that self-trapping occurs when $|g| > 4$. Nonetheless, we now have a proof that self-trapping occurs when $|g|$ is sufficiently large.

Put another way, it is known numerically that $|g_c| = 4$, and while a proof is not known, we have proved an upper bound that $|g_c| \le 7.22$ when starting at the end of a path, and $|g_c| \le 9$ when starting in the interior of a path or anywhere on a cycle. That is, $|g_c| \le f_\text{min}$.

For a given $|g| > f_\text{min}$, we can use $f(p_+) = |g|$ to find $p_+$, which is a lower bound on the self-trapping probability with that value of $|g|$. For example, for a quantum walk starting at an endpoint of a path, say we have $|g| = 20$. Solving $f(p_+) = |g|$ with $\operatorname{deg}(v_r) = 1$ and $|g| = 20$ numerically yields $p_+ \approx 0.987$, so when $|g| = 20$, the probability that the walker stays at its initial vertex is at least $0.987$ for all time.

Conversely, if we desire that the self-trapping probability be at least some level, that is equivalent to choosing $p_+$, from which we can find $|g|$ using $|g| = f(p_+)$. For example, say we want a quantum walk to stay at its initial vertex on a cycle with at least $0.95$ probability. Evaluating $|g| = f(p_+)$ with $\operatorname{deg}(v_r) = 2$ and $p_+ = 0.95$ yields $|g| \approx 15.1$, so as long as we choose $|g|$ to be at least 15.1, the probability of self-trapping will be at least $0.95$.


\section{\label{sec:transfer}Timed Quantum State Transfer and Quantum Memory}

\begin{figure}
\begin{center}
    \subfloat[] {
        \includegraphics{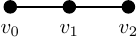}
        \label{fig:path-3}
    } \qquad\qquad\qquad
    \subfloat[] {
        \includegraphics[width=7.2cm]{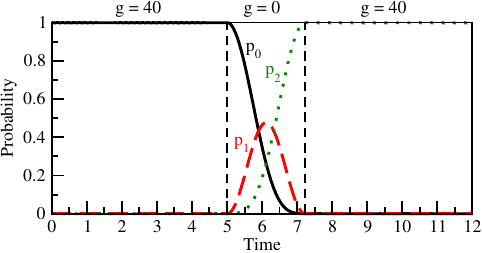}
        \label{fig:pst-buffer}
    }
    \caption{(a) The path graph $P_3$. (b) With the walker initially at $v_0$, the probability at each vertex of $P_3$, labeled $p_0$, $p_1$, and $p_2$, as a function of time. The vertical dashed lines separate times when different values of $g$ were used: $g = 40$ on the left, $g = 0$ in the middle, and $g = 40$ again on the right.}
\end{center}
\end{figure}

Quantum state transfer is the task of moving a qubit from one location to another, and in a spin chain, this is precisely a quantum walk \cite{Christandl2005}. The walker starts at a source vertex, and the goal is for the walker to end up at a target vertex at some time with a high probability, or fidelity. When the fidelity is 1, it is called perfect state transfer \cite{Christandl2005}, and when the fidelity can be made arbitrarily close to 1, it is called pretty good state transfer \cite{Godsil2012}. Experimentally, quantum state transfer has been demonstrated using photonic waveguide lattices \cite{Chapman2016}, via nuclear magnetic resonance \cite{Singh2023}, and with superconducting qubits \cite{Li2018,Shang2020,Ge2026}.

Now, we propose using self-trapping in nonlinear quantum walks to control the timing of state transfer. This is done by choosing a sufficiently large value of $|g|$ to trap the walker in place. Then, when the transfer is ready to begin, we can set $g = 0$ to perform the transfer. When the transfer is completed, we can set $|g|$ to a large value again to trap it until the quantum information is ready to be used. 
As mentioned in the introduction, it is possible to dynamically change $g$ in some physical systems, such as Bose-Einstein condensates \cite{Timmermans1999} and in nonlinear waveguide arrays \cite{Cui2022}.

For example, consider the path of three vertices, $P_3$, shown in \fref{fig:path-3}. It is known to support perfect state transfer from one end of the path to the other end, i.e., between $v_0$ and $v_2$ in time $\pi/\sqrt{2}$ \cite{Christandl2005,Godsil2012}. In \fref{fig:pst-buffer}, we plot the probability in each of the three vertices of $P_3$, and the Mathematica code to generate it is available in \cite{data}. The walker starts at $v_0$. From the start until $t = 5$, we use $g = 40$, and \fref{fig:pst-buffer} shows that in this region of time, $p_0$ is near 1, while $p_1$ and $p_2$ are near 0. In fact, from our result from the last section, solving $f(p_+) = |g|$ yields a self-trapping probability of at least $0.997$. At $t = 5$, we want to start the transfer, so we change the nonlinearity coefficient to $g = 0$, and the walker moves to the other end of $P_3$ in time $\pi/\sqrt{2} \approx 2.22$. In \fref{fig:pst-buffer}, we see that $p_0$ drops and $p_2$ rises during this interval. Finally, at $t = 7.22$, we change $g$ back to $40$, trapping the walker at $v_2$.

Another potential interpretation of this holding and transmitting of a qubit is a quantum memory. Although quantum memories typically use one type of qubit as storage qubits and another type for flying qubits, in our case, the same qubit plays both roles, and we use trapping and releasing via state transfer to determine whether it is in storage or flying mode.


\section{\label{sec:conclusion}Conclusion}

In this work, we studied the behavior of continuous-time quantum walks on general path and cycle graphs governed by the cubic nonlinear Schr\"odinger equation. While it was previously numerically known that self-trapping occurs when the nonlinearity coefficient $g$ is of sufficiently high magnitude, we have analytically derived values for $|g|$ beyond which self-trapping is guaranteed to occur, and it only depends on the degree of the initial vertex, not the number of vertices in the lattice. There exist values of $g$ for which self-trapping occurs, but which lie beyond the conditions for our proof, and proving that self-trapping occurs in this regime remains an open problem.

In addition, our analysis gives a quantitative relation between the nonlinearity coefficient $g$ and the self-trapping probability. For any given value of $g$, one can determine a corresponding lower bound on the self-trapping probability, and for any given self-trapping probability, one can determine a corresponding lower bound on $|g|$.

As a potential application, by turning on and off self-trapping by adjusting $|g|$, we can control the timing of quantum state transfer, which is equivalent to storing and releasing quantum information, i.e., quantum memory.


\ack{This material is based upon work supported in part by the National Science Foundation EPSCoR Cooperative Agreement OIA-2044049, Nebraska’s EQUATE collaboration. Any opinions, findings, and conclusions or recommendations expressed in this material are those of the author(s) and do not necessarily reflect the views of the National Science Foundation.}

\data{The data that support the findings of this article are openly available \cite{data}.}

\roles{
Yujia Shi \orcid{0000-0002-4044-5659} Conceptualization (equal), Investigation (lead), Methodology (lead), Validation (equal), Writing – original draft (lead), Writing – review \& editing (equal)

Thomas G.~Wong \orcid{0000-0002-5019-7302} Conceptualization (equal), Funding acquisition, Investigation (supporting), Methodology (supporting), Project administration, Supervision, Validation (equal), Writing – original draft (supporting), Writing – review \& editing (equal)
}


\appendix

\section{\label{appendix}Proof of Inequality Relating Nonlinearity Coefficient and Self-Trapping Probability}

In this appendix, we prove \eqref{eq:bound-g-p}, which relates the nonlinearity coefficient $|g|$ to the self-trapping probability.

Assume $g\neq0$. Without loss of generality, we can permute the labels of the vertices so that the first component of $\ket{\psi(t)}$ is the amplitude at the initial vertex $v_r$, which lets us write $\ket{\psi(t)}$ in the following block form:
\[
    \ket{\psi(t)} = \begin{pmatrix} a(t) \\ \ket{b(t)} \end{pmatrix},
\]
where $a(t)=\psi_r(t)\in\mathbb{C}$ is a scalar that is the amplitude at the initial vertex $v_r$, and $\ket{b(t)} \in\mathbb{C}^{N-1}$ is a sub-vector that contains the amplitudes at the remaining vertices. We denote the probability at the initial vertex $v_r$ by
\[
p(t) = |\psi_r(t)|^2 = |a(t)|^2,
\]
and so the magnitude of the amplitude at the initial vertex is
\[ |a(t)| = \sqrt{p(t)}. \]
Since the probability of finding the walker at any vertex is 1, we also have
\begin{align*}
    \braket{b(t)}{b(t)} 
        &= \sum_{j \ne r} |\psi_j(t)|^2 = 1 - |\psi_r(t)|^2 = 1 - |a(t)|^2 \\
        &= 1 - p(t),
\end{align*}
and so the $L^2$-norm of $\ket{b(t)}$ is
\[ \| \ket{b(t)} \| = \sqrt{1-p(t)}. \]
With respect to this decomposition, the Hamiltonian takes the block form
\[
H=-A =
\begin{pmatrix}
0 & -\bra{h} \\
-\ket{h} & -B \\
\end{pmatrix},
\]
where $\ket{h} \in \mathbb{C}^{N-1}$ is a real-valued column vector of length $N-1$ that contains the edges from $v_r$ to its neighbors, and $B \in \mathbb{C}^{(N-1)\times(N-1)}$ is the adjacency matrix on the remaining vertices. Note $\| \ket{h} \| = \sqrt{\operatorname{deg}(v_r)}$, so if $v_r$ is an endpoint of a path, it has one neighbor and $\|h\|_2 = 1$, and if is an interior vertex of a path or any vertex on a cycle, it has two neighbors and $\|h\|_2 = \sqrt{2}$. $\|B\|=\sup_{\|x\|_2=1}\|Bx\|_2$ denotes the operator norm induced by the Euclidean norm. It is equal to the largest absolute value of its eigenvalues since $B$ is a symmetric matrix. Removing $v_r$ from a path leaves either one path or a disjoint union of two paths. Since the adjacency matrix of any path has spectral norm at most $2$, and the norm of a block diagonal matrix is the maximum of the norms of its blocks, in both cases, we have $\|B\|\le 2$.

Next,
\begin{align*}
\langle \psi| H |\psi \rangle
&= \begin{pmatrix} 
    a^* & \bra{b} \\
\end{pmatrix} \begin{pmatrix} 
    0 & -\bra{h} \\
    -\ket{h} & -B \\
\end{pmatrix}
\begin{pmatrix} 
    a \\ 
    \ket{b} \\
\end{pmatrix} \\
&= \begin{pmatrix} 
    a^* & \bra{b} 
\end{pmatrix} \begin{pmatrix} 
    -\braket{h}{b} \\
    -a\ket{h} - B\ket{b} \\
\end{pmatrix} \\
&=
-a^* \braket{h}{b} - a \braket{b}{h} - \langle b | B | b \rangle \\
&=
-2 \operatorname{Re}(a \braket{b}{h}) - \langle b | B | b \rangle.
\end{align*}
Taking the magnitude,
\[ | \langle \psi| H |\psi \rangle | = \left| -2 \operatorname{Re}(a \braket{b}{h}) - \langle b | B | b \rangle \right|, \]
and from the triangle inequality,
\[ | \langle \psi| H |\psi \rangle | \le 2 \left|\operatorname{Re}(a \braket{b}{h})\right| + \left|\langle b | B | b \rangle \right|. \]
For the first term,
\begin{align*}
2 \left|\operatorname{Re}(a \braket{b}{h})\right|
&\le 2 |a|\, |\braket{b}{h}| \\
&\le 2 |a|\, \| \ket{b} \|\, \| \ket{h} \| \\
&= 2\, \sqrt{p} \sqrt{(1-p)} \sqrt{\operatorname{deg}(v_r)} \\
&= 2\, \sqrt{p(1-p)\operatorname{deg}(v_r)},
\end{align*}
where in the second line, we used the Cauchy-Schwarz inequality. For the second term,
\[
\left|\langle b | B | b \rangle \right| \le \|B\|\, \braket{b}{b} \le 2(1-p),
\]
where the first inequality holds because $B$ is a real-valued adjacency matrix, so it is Hermitian, and so $\langle b | B | b \rangle$ is real.
Plugging in these bounds,
\[
\left|\langle \psi | H |\psi \rangle\right| 
\le
2\, \sqrt{p(1-p)\operatorname{deg}(v_r)} + 2(1-p).
\]
Since $H$ is Hermitian, $\langle \psi | H | \psi \rangle$ is real, so it is bounded above by this bound and below by the negative of the bound, i.e.,
\begin{equation}\label{ineq:1_appendix}
-2\, \sqrt{p(1-p)\operatorname{deg}(v_r)} - 2(1-p)
\le
\langle \psi | H |\psi \rangle \nonumber
\le
2\, \sqrt{p(1-p)\operatorname{deg}(v_r)} + 2(1-p).
\end{equation}

Now, it is known \cite{Wong44,Brazhnyi2013} that the Gross-Pitaevskii equation causes the state $\ket{\psi(t)}$ to evolve in such a way that the expected value of
\[ H - \frac{1}{2} g |\psi|^2 \]
is conserved. Explicitly taking the expected value by sandwiching between $\bra{\psi}$ and $\ket{\psi}$, the quantity
\begin{align*}
    E(\psi) 
        &= \left\langle \psi \middle | H - \frac{1}{2} g |\psi|^2 \middle| \psi \right\rangle \\
        &= \langle \psi | H | \psi \rangle - \frac{1}{2} g |\psi(t)|^2 \braket{\psi}{\psi} \\
        &= \langle \psi | H | \psi \rangle - \frac{1}{2} g |\psi(t)|^4
\end{align*}
is conserved. Evaluating this at $t = 0$ using $\ket{\psi(0)} = \ket{r}$, we get
\[
E(\psi(0)) = -\frac{g}{2}.
\]
Hence, for all $t$,
\[
    -\frac{g}{2}
=
E(\psi(t))
=
\langle \psi | H |\psi \rangle - \frac{g}{2} \sum_j |\psi_j|^4,
\]
or rearranging,
\begin{equation}\label{eq:conservedenergy_appendix}
\langle \psi | H |\psi \rangle = -\frac{g}{2} + \frac{g}{2} \sum_j |\psi_j|^4.
\end{equation}
Moreover,
\begin{align}
\sum_j |\psi_j|^4
&=
|a|^4 + \sum_{j \ne r} |\psi_j|^4 \nonumber \\
&\le |a|^4 + \sum_{j \ne r} |\psi_j|^2 \sum_{k \ne r} |\psi_k|^2 \nonumber \\
&= p^2 + (1-p)^2. \label{ineq:sumpsi_appendix}
\end{align}
Combining \eqref{eq:conservedenergy_appendix} and \eqref{ineq:sumpsi_appendix}, when $g > 0$, we obtain an upper bound on $\langle \psi | H | \psi \rangle$,
\[ \langle \psi | H | \psi \rangle \le -\frac{g}{2} + \frac{g}{2} \left[ p^2 + (1-p)^2 \right] = -|g|p(1-p), \]
and when $g < 0$, we obtain a lower bound on $\langle \psi | H | \psi \rangle$,
\[ \langle \psi | H | \psi \rangle \ge -\frac{g}{2} + \frac{g}{2} \left[ p^2 + (1-p)^2 \right] = |g|p(1-p). \]

When $g > 0$, we can combine the above upper bound with the lower bound from \eqref{ineq:1_appendix}. Similarly, when $g < 0$, we can combine the above lower bound with the upper bound from \eqref{ineq:1_appendix}. In both cases, we get
\[
|g|\,p(1-p)
\le
2\sqrt{p(1-p)\operatorname{deg}(v_r)}
+2(1-p).
\]
Solving for $|g|$ and assuming $0 < p < 1$, we obtain
\[
    |g| \le 
\frac{2 \sqrt{\operatorname{deg}(v_r)}}{\sqrt{p(1-p)}}
+
\frac{2}{p}.
\]
Swapping the two sides, we get the inequality as reported in the main text.


\bibliographystyle{iop}
\bibliography{refs}

\end{document}